\documentclass[a4paper,12pt]{article}

\usepackage{amsmath}
\usepackage{graphicx}
\usepackage{bm}
\usepackage{subfigure}
\usepackage[margin=15pt,font=small,labelfont=bf]{caption}
\newcommand{\mrm}{\mathrm}

\begin{document}

\title{A source of ultra-cold neutrons for the gravitational spectrometer "GRANIT"}
\author{P. Schmidt-Wellenburg\thanks{\textit{Corresponding author: Tel.: +33
4 76 20 70 27; fax: +33 4 76 20 77 77. E-mail address: schmidt-w@ill.fr}}~$%
^{1,2}$, P. Geltenbort$^1$, V.V. Nesvizhevsky$^1$,\\ C. Plonka$^1$, T. Soldner$^1$, F. Vezzu$^3$, O. Zimmer$^{1,2}$\\
\textit{\small 1) Institut Laue Langevin,}\\
\medskip \textit{\small 6, rue Jules Horowitz, BP-156, 38042 Grenoble
Cedex 9, France}\\
\textit{\small 2) Physik-Department E18,}\\
\medskip \textit{{\small Technische Universit\"{a}t M\"{u}nchen, 85748 Garching,
Germany}}\\
\textit{\small 3) Laboratoire de Physique Subatomique et de Cosmologie,}\\
\textit{{\small IN2P3/4JF, 53 Avenue des Martyrs, 38026 Grenoble Cedex,
France}}\\
}
\maketitle

\abstract{We present the status of the development of a dedicated high density ultra-cold neutron (UCN) source dedicated to the gravitational spectrometer GRANIT. The source employs superthermal conversion of cold neutrons to UCN in superfluid helium. Tests have shown that UCN produced inside the liquid can be extracted into vacuum. Furthermore a dedicated neutron selection channel was tested to maintain high initial density and extract only neutrons with a vertical velocity component $v_{\bot}\leq 20$~cm/s for the spectrometer. This new source would have a phase-space density of $\Gamma_{\mrm{\scriptscriptstyle{UCN}}} \approx 0.18~\mrm{cm^{-3}(m/s)^{-3}}$ for the spectrometer.}\\

\section*{Introduction}
The solutions for Schr\"{o}dinger's equation for a neutron bouncing on a reflecting horizontal surface in the earth gravitational field are given by Airy functions\,\cite{Luschikov1978}. This textbook example of bound energy states in a linear potential has been demonstrated experimentally at the high flux reactor of the Institut Laue Langevin\,\cite{Nesvizhevsky2002,Nesvizhevsky2003}. A new gravitational spectrometer, GRANIT\,\cite{Nesvizhevsky}, is being built to investigate these quantum states further, and to induce resonant transitions between gravitationally bound quantum states. Applications are a refined measurement of the electrical charge of the neutron, the search for the axion\,\cite{Baessler2007}, and other additional forces beyond the standard model.
Experiments at the present UCN source are limited by counting statistics and systematic effects\footnote[1]{For a discussion of systematic effects concerning these measurements see G.~Pignol (these proceedings).}. In these conference proceedings the advances of a dedicated UCN source, based on superthermal conversion of cold neutrons in superfluid helium are described.

\section*{General aspects of Helium-4 based superthermal converters}
The dominant process in the conversion is the excitation of a single phonon inside the superfluid helium. Cold neutrons with a wavelength around 8.9~\AA\, ($k=2\pi/\lambda^{\ast}=0.7$~\AA$^{-1}$), i.e.\,1.0~meV kinetic energy, can be scattered down to the ultra-cold energy range by emission of one single phonon\,\cite{Golub1975}. Multiphonon processes may also occur, depending on the incoming neutron spectrum\,\cite{Korobkina2003,Schott2003}. The resulting saturated UCN density,
\begin{equation}
	\rho_{\mathrm{\scriptscriptstyle{UCN}}}=P \cdot \tau,	
\end{equation}
\label{eq:density}

\noindent is determined by production rate density $P$ and neutron storage time $\tau$ inside the converter volume. The rate
\begin{equation}
	1/\tau = \frac{1}{\tau_{\beta}}+\frac{1}{\tau_{\mrm{wall}}}+\frac{1}{\tau_{\mrm{up}}}+\frac{1}{\tau_{\mrm{abs}}},
\end{equation}
\label{eq:lifetime}

\noindent includes contributions from all existing loss channels (neutron beta decay, losses due to wall collisions, up-scattering due to phonons, nuclear absorption by $ ^3$He\,-\,impurities in the helium). As the absorption cross section of $ ^4$He is zero there is no absorption inside a pure converter.\\
The equation of detailed balance,
\begin{equation}
	\sigma\left(E_{\mathrm{\scriptscriptstyle{UCN}}} \longrightarrow E^{\ast} \right) = \frac{E^{\ast}}{E_{\mathrm{\scriptscriptstyle{UCN}}}}\exp{\left(-\frac{E^{\ast}-E_{\mathrm{\scriptscriptstyle{UCN}}}}{k_{\mrm{B}}T}\right)} \cdot \sigma\left( E^{\ast} \longrightarrow E_{\mathrm{\scriptscriptstyle{UCN}}} \right),
\end{equation}
\noindent relates the cross sections of a two level system, as our one-phonon excitation process, dependent on temperature. For $T \longrightarrow 0$~K the up-scattering cross section will be arbitrarily small. In pure $ ^4$He only two contributions to the total storage time remain; the neutron beta decay and losses due to wall collisions. The UCN production rate density is defined as the conversion rate of neutrons to energies below the Fermi potential of the walls of the converter volume and is thus dependent on the wall material. The production rate density expected for the 1-phonon process in a beryllium ($V_{\mrm{F}}=252$~neV) coated converter volume~($V_{\mrm{F}}=252$~neV) is $P_{\mrm{I}}=(4.55\pm 0.25) \cdot 10^{-8} \mrm{d}\Phi/\mrm{d}\lambda|_{\lambda^{\ast}} \mrm{s^{-1}cm^{-3}}$, with the differential flux at $\lambda^{\ast} = 8.9$~\AA~given in $\mrm{cm^{-2}s^{-1}}$\,\AA\,$^{-1}$\,\cite{Baker2003}.

\section*{Source concept}
\begin{figure}
	\centering
	\includegraphics[width=0.8\textwidth]{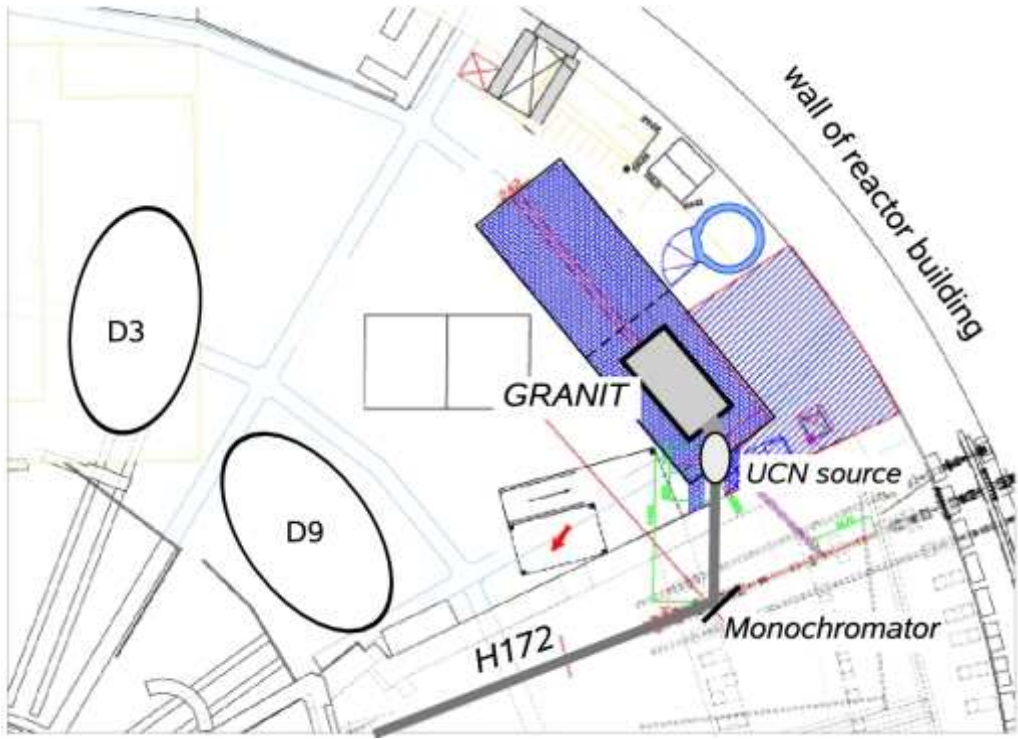}
	\caption{Setting of the monochromator, the source, and GRANIT inside level C of the ILL. (For more details see: www.ill.fr/pages/science/imgs/PlanInstILL.gif)}
        \label{fig:position}
\end{figure}

GRANIT and its source will be positioned on the neutron beam H172 on level C of the high flux reactor at the ILL\,(see fig.\,\ref{fig:position}). A crystal monochromator, positioned 12~m downstream from the cold source, will reflect 8.9~\AA\, neutrons under a take-off angle of $\theta = 61.2^{\circ}$. Between monochromator and converter volume a 3~m long neutron guide with $m=2$ supermirror coating will be installed. The reduction in background will outweight by far the decrease in conversion rate due to the imperfect monochromator and the omission of multi-phonon processes. A nitrogen cooled beryllium filter will be installed further upstream of the monochromator to reduce high energy neutrons at the monochromator position. As source we will employ an existing apparatus\,\cite{Zimmer2007} modified for our new requirements; a continuous flux of UCN within a narrow phase space element. 

\begin{figure}
	\centering
	\includegraphics[width=0.6\textwidth]{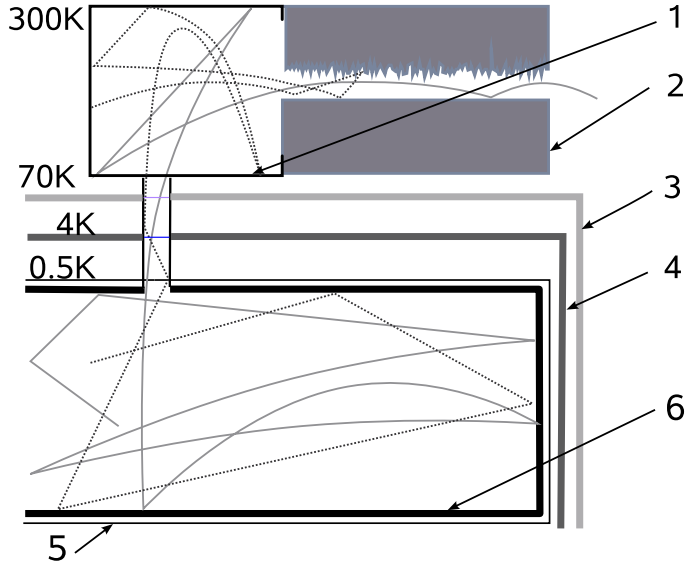}
	\caption{Schematic sketch of the source with extraction. 1)\,Intermediate storage volume, 2)\,semidiffuse extraction channel, 3)\,70~K heat screen, 4)\,4~K heat screen, 5)\,vessel for superfluid helium, 6)\,converter volume. Neutrons entering the channel with  $E_{\bot}\geq mgh$\,(-~-~-) are reflected back, the energy components are redistributed due to the rough surfaces of the intermediate volume thus it may pass the channel in the next attempt. A neutron with $E_{\bot}\leq mgh$\,(---) can pass the channel immediately. }
	\label{fig:extraction}
\end{figure}

A sketch of the converter volume with heat screens and extraction is shown in figure\,\ref{fig:extraction}.
The UCN will diffuse from the converter volume through a highly specular guide into an intermediate volume. This second volume at room temperature allows a vertical tube extraction from the converter. Thus avoiding cryogenic difficulties of a direct horizontal extraction with foils which has posed problems in the past\,\cite{Kilvington1987} and reduces background from directly scattered cold neutrons. Furthermore it evenly distributes all neutrons to the semidiffuse channel\,\cite{Schmidt-Wellenburg2007}. Which is used to select neutrons with small vertical energy $E<mgh<20$~peV for the spectrometer whereas all other neutrons are reflected back or are lost on wall collisions. This method will increase the storage time inside the intermediate volume and the converter volume, as more than 85\% of the neutrons incident on the channel entrance will be reflected back. Thus the UCN density is increased which in turn increases the flux of extracted neutrons.

\subsection*{Neutron monochromator}
Only a narrow range of wavelengths around 8.9~\AA\, of incident neutrons is suitable for the dominant 1-phonon downscattering in superfluid helium. This allows to benefit from a crystal monochromator which reflects apt neutrons and diminishes background at the experiment.\\

\begin{figure}
	\centering
	\includegraphics[width=0.8\textwidth]{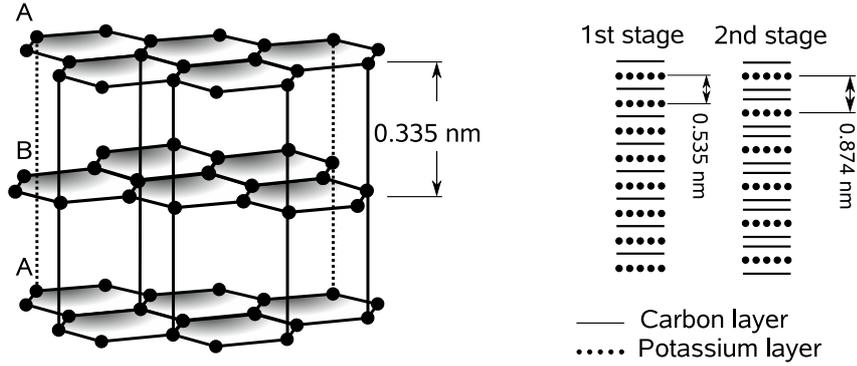}
	\caption{Crystal structure of graphite and staging of GiC. The intercalant diffuses into the graphite and thus increases the lattice spacing. Different stages of GiC can be produced. The stage number refers to the number of unperturbed graphite layer between two layers of intercalant atoms.}
	\label{fig:gic}
\end{figure}

Crystal monochromators reflect neutrons of the desired wavelengths away from the primary beam under the $d$-spacing dependent Bragg angle:
\begin{equation}
	\theta_{\mrm{B}}=\arcsin{\left(\frac{n\lambda}{2d}\right)}.
\end{equation}
The $d$-spacing of a monochromator for $\lambda=8.9$~\AA\, has to be $d>\lambda/2 = 4.45$~\AA. In a perfect single crystal the line width for Bragg reflected neutrons is extremely small, $\Delta k/k = 10^{-4}$\,\cite{Goldberger1947} leading to a very narrow acceptance angle of incident neutrons:
\begin{equation}
	\frac{\Delta k}{k} = \cot{\theta \Delta \theta}.
\end{equation}
The incident beam has a divergence of typically $\pm 2^{\circ}$ at $8.9$~\AA\, due to the $m=2$ supermirrors used in the guide. Therefore a "mosaic crystal"\,\cite{Goldberger1947} will be used. Such a crystal can be regarded as a collection of microscopic small perfect crystals with differing angles $\epsilon$ with respect to the overall crystal orientation. Although the angular distribution is in general arbitrary, it resembles a cylindrically symmetric Gaussian distribution:
\begin{equation}
	W(\epsilon) = \frac{1}{\sqrt{2 \pi} \eta} \exp{\left( -\epsilon^2 /2 \eta^2 \right)},
\end{equation}

\noindent where $\eta$, called the mosaicity, is the width. To obtain a high acceptance the mosaicity should be in the range of the divergence of the incident beam\,\cite{Liss1994}. A high integral reflectivity of the crystal is demanded to ensure an acceptable $8.9$~\AA\, flux for the converter. This requires a small absorption cross-section.\\
There are two materials which match these requirements, mica and graphite intercalated compounds\,(GiC) with alkali metals. Due to the small mosaicity $\approx 0.3^{\circ}$ of mica we are employing a potassium intercalated graphite monochromator of the type previously developed at the NIST\,\cite{Mattoni2004}.  
The d-spacing for highly oriented pyrolytic graphie is $d=3.35$~\AA\, with a typical mosaicity of $1^{\circ} - 2^{\circ}$. It is increased by placing guest species in between graphite layers (see fig.\,\ref{fig:gic}), this process is called intercalation.\\

The monochromator for GRANIT will be made of stage-2 ($\mrm{C_{24}K}$) potassium intercalated graphite providing a lattice spacing of $d=8.74$~\AA\, giving a take-off angle of $2\theta = 61.2^{\circ}$. The typical mosaic spread of the produced crystals is $\eta = 1.5^{\circ} - 2.2^{\circ}$ which matches fine with the incident divergence of beam H172. An integral reflectivity of $r\geq 80$~\% can be achieved\,\cite{Mattoni2004}. Furthermore the absorption cross section of graphite (0.0035~barn) and potassium (2.1~barn) are rather small. The $d$-spacing of Rubidium and Caesium intercalated compounds are of the same magnitude. Rubidium would be an interesting alternative due to its small cross section of 0.38~barn, whereas Cesium (29~barn) is less interesting. For these two alkali metals the production method is not yet sufficiently refined.

\begin{figure}
	\centering
	\includegraphics[width=0.8\textwidth]{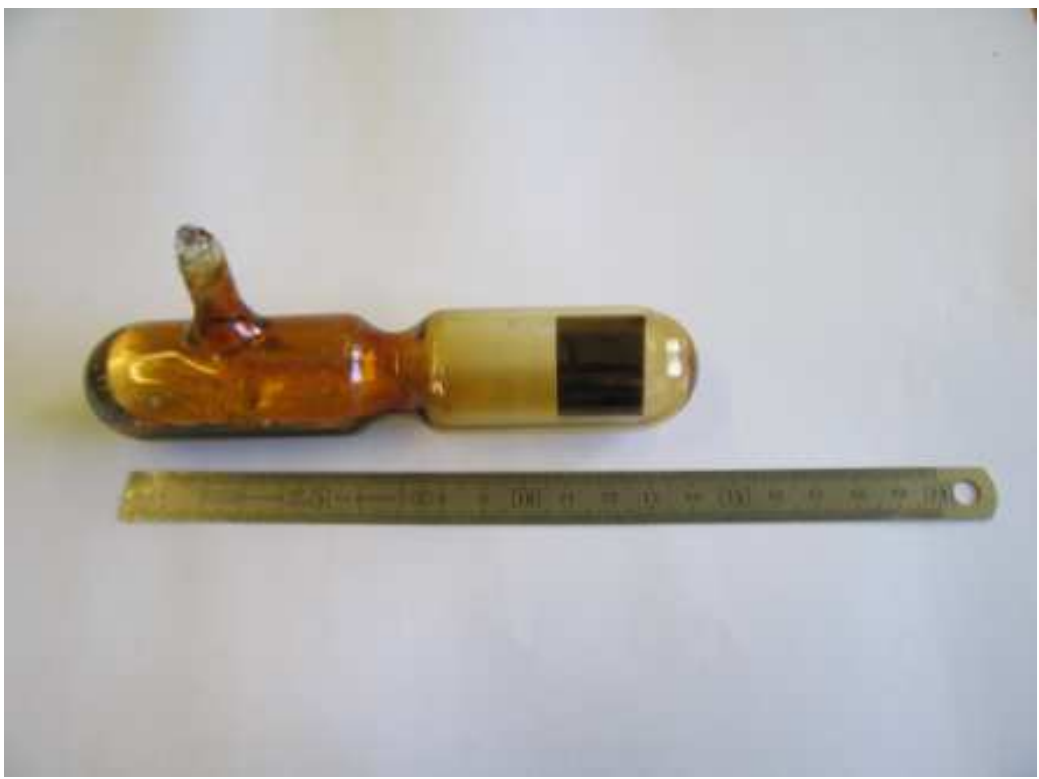}
	\caption{Typical cell made from Pyrex used for intercalation. The potassium is on the left and causes at $T\geq 250~^{\circ}$C the brown colouring of the glass. On the right: a stage-2 intercalated graphite.}
	\label{fig:two-bulb}
\end{figure}

Alkali intercalated graphite compounds are conveniently produced using the "two-bulb" technique, where the graphite is maintained at a temperature $T_{\mrm{c}}$ which is higher than $T_{\mrm{a}}$ of the alkali metal\,\cite{Herold1955}. The stage which is formed depends on the temperature difference $\Delta T =  T_{\mrm{c}}-T_{\mrm{a}}$ and the quantity of potassium available. For stage-1, stage-2 we are employing 5\,g, 1\,g~ampoules of potassium at $T_{\mrm{a}}=255\pm 3~^{\circ}\mrm{C}$ and $\Delta T_1 = 10\pm 3~^{\circ}\mrm{C}$, $\Delta T_2 = 102\pm 3~^{\circ}\mrm{C}$, respectively. A photograph of such a two-bulb cell is shown in figure\,\ref{fig:two-bulb}.

\begin{figure}
	\centering
	\includegraphics[width=0.8\textwidth]{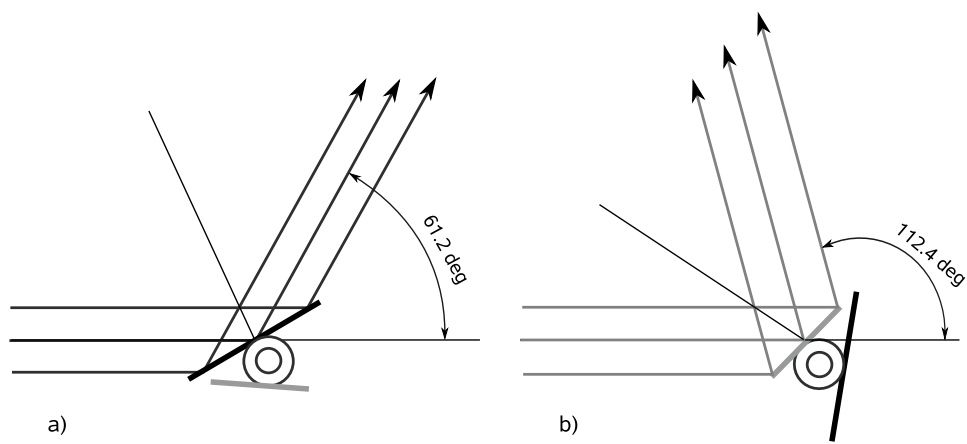}
	\caption{Monochromator setup on rotary table. a)\,stage-2 potassium intercalated graphite monochromator in position, b)\,stage-1 (potassium) monochromator. This setup allows relatively short switching time.}
	\label{fig:mono}
\end{figure}

A second monochromator with a take-off angle of $2\theta = 112.5^{\circ}$ will include a set of stage-1 (C$_{8}$K) crystals ($d=5.35$~\AA) placed close to the first on a rotary table. The two different monochromators can be interchanged (see fig.\,\ref{fig:mono}) making two separate $8.9$~\AA~ beam ports at H172 available. The second beam will feed a position with neutrons for further tests and developments on liquid helium based UCN-sources, and later the cryo-EDM experiment\,\cite{Baker2003}. 

\subsection*{Converter and cryostat}
As the incident neutron beam has a divergence of $\alpha \approx 2^{\circ}$, the differential flux will be a function of position inside the volume. In order to guide divergent cold neutrons incident on the walls it is convenient to use polished wall materials with a high critical angle. For the storage of UCN we need a high Fermi potential, rough surfaces for an isotrop momentum distribution, and a small loss per bounce coefficient $\mu$ to minimise wall losses. Obviously it is impossible to have polished and rough surfaces at the same place, therefore the source will have rough surfaces only on the entrance and exit window of the converter volume, whereas the rest is polished. Beryllium, beryllium-oxide, and diamond like carbon\,(DLC) all have Fermi potentials $V_{\mrm{F}}\geq 250$~neV, and $\mu \leq 1\cdot10^{-4}$.\\

\begin{figure}
	\centering
	\includegraphics[width=0.9\textwidth]{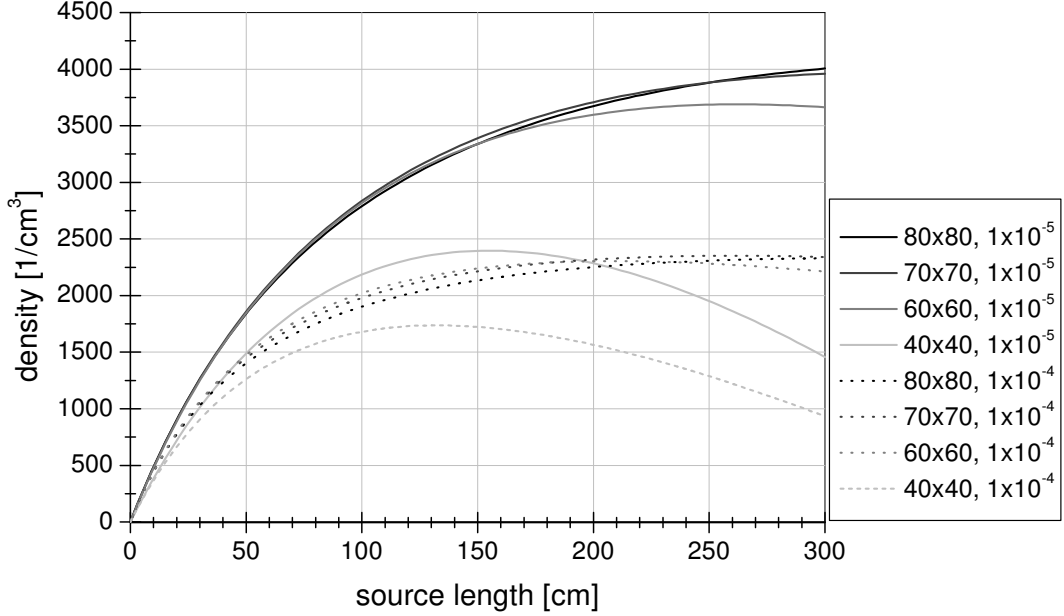}
	\caption{Simulation of UCN-density as a function of converter volume length, various rectangular cross sections, and two wall loss coefficients. The incident beam is taken to have a Gaussian divergence $\alpha \geq 1.8^\circ$ dependent on the converging guide and differential flux $\mrm{d}\phi/\mrm{d}\lambda |_{\lambda^{\ast}}=3\cdot10^8~\mrm{cm^{-2}s^{-1}}$\,\AA\,$^{-1}$ at the entrance of the volume. These values result from the simulations described in the text. The extraction opening is a hole of {\O} = 10~mm. UCN once passed through the hole are taken to be lost from the volume, in reality there is a certain reflection probability.} 
		\label{fig:density_sim}
\end{figure}

Calculations for a divergent beam, a Fermi potential $ V_{\mrm{F}} = 252$~neV, a mirror like wall ($m=1$), and varying loss coefficients are shown in figure\,\ref{fig:density_sim}. The incident beam flux on the monochromator with a reflectivity of $r=85$~\%,  was taken to be $\mrm{d}\phi/\mrm{d}\lambda|_{\lambda^{\ast}} = 6\cdot 10^8 \mrm{s^{-1}cm^{-3}}$. This value was calculated from known cold source data and a transmission simulation for the existing guide. The $m=2$ neutron guide between monochromator and source converges from a $80\times 80 \mrm{mm^2}$ to the cross section indicated in the figure over a length of $3$~m. Both guide parts have been simulated with a Monte~Carlo algorithm. 

The combination of simulated transmissions and divergence for the beam and analytic calculations for a converter volume of length 600~mm, a cross section of 70$\times$70~mm$^2$, and an average $\mu=1\cdot10^{-4}$ give a density of $\rho_{\scriptscriptstyle{calc}}\geq 1250~\mrm{cm^3}$.

The behaviour of the coatings during cooldown still has to be investigated. Thermal tension could create cracks and degrade the surface quality significantly. A cooling test with several coatings and substrates will be performed to find the best choice. 
For filling and cooling the converter volume with liquid helium we employ a cryostat previously developed at Munich\,\cite{Zimmer2007,Schmidt-Wellenburg2006}. Primary cooling power is provided by a Gifford McMahon cryocooler with 1.5~W at 4.2~K. It liquefies helium and provides the cooling power for the 50~K and 4~K heat screens. The liquefaction is described in detail in ref.\,\cite{Schmidt-Wellenburg2006}. A $ ^4$He evaporation stage cools the liquid helium below the $\lambda$-transition point $T_{\lambda}=2.177$~K, which then allows us to use a superleak to remove $ ^3$He. The cooling of the converter volume is achieved with a $ ^3$He closed cycle evaporation stage. Using a roots blower pump with 500~$\mrm{m^3/h}$ nominal pumping speed backed by a 40~$\mrm{m^3/h}$ multiroots pump we were so far able to cool the filled converter to 0.7~K. Further improvements, including a new heat exchanger between $ ^3$He and $ ^4$He will allow us to reach temperatures below 0.5~K. More details about the cryostat will be published elsewhere.

\subsection*{UCN selection with semidiffuse channel}
The converter volume will be connected to the intermediate volume by a highly polished, beryllium, nickel, or DLC coated guide. This additional volume will be made of DLC coated aluminium plates. Having rough surfaces it distributes the UCN evenly on the entrance window of the semidiffuse channel and avoids cryogenic complications which would arise form a direct attachement of the channel to the converter volume. The optimum cross section was determined by simulations with Geant4UCN\,\cite{atchinson2005} to be 40$\times$40~$\mrm{mm^2}$, the length 300~mm is given by the dimensions of the spectrometer. A challenging task is to minimise heat input without reflecting UCN back into the converter volume before they have reached the intermediate volume. Using thin aluminium or mylar foils as heat screens seems to be an option, although losses due to multiple passages will increase due to multiple passages.
 The UCN will be extracted via a 0.2~mm high horizontal semidiffuse extraction channel (see fig.\,\ref{fig:extraction}). Measurements\,\cite{Schmidt-Wellenburg2007} have shown that such a channel increases the storage time and selects only the desired fraction of the phase space. The extraction channel is made of DLC coated quartz plates\,\cite{Nesvizhevsky2007}, the lower surfaces are polished, the upper rough. The channel dimensions are $h=200~\mrm{\mu m},\,l=100~$mm, $w\,=\,300~$mm, which will provide a reflectivity $r\geq 85$~\% for neutrons with a vertical energy component $E_{\bot}\geq mgh = 20$~peV.

\section*{Conclusion}
The crystals of the monochromator, the converter, and the semidiffuse channel have all been tested separately. Their integration remains a challenging task. Present calculation with an incident differential flux $\mrm{d}\phi/\mrm{d}\lambda|_{\lambda^{\ast}} = 6\cdot 10^8~\mrm{s^{-1}cm^{-3}}$ on the monochromator give a UCN density of $\rho_{\mrm{\scriptscriptstyle{UCN}}} \approx 1250~\mrm{cm^{-3}}$ in the converter, $\rho_{\mrm{\scriptscriptstyle{int}}} = 800~\mrm{cm^{-3}}$ in the intermediate volume. This yields for a critical velocity of $7$~m/s for the materials used an available phase-space density of $\Gamma_{\mrm{\scriptscriptstyle{He}}} \approx 0.18~\mrm{cm^{-3}(m/s)^{-3}}$. Compared to the phase-space density $\Gamma_{\mrm{\scriptscriptstyle{Turbine}}} \approx 0.013~\mrm{cm^{-3}(m/s)^{-3}}$ of the UCN turbine at the ILL this is more than a factor ten of improvement. This calculation assumes perfect conditions and in all parts optimal transmissions, experience with other simulations have shown that the actual value might be lower.

\section*{Acknowledgement}
We are greatful to all our colleagues from the GRANIT collaboration, the monochromator collaboration with NIST, and the DPT of the ILL for fruitful discussions and support. This work is partly supported by the French Agence de la Recherche (ANR)

\end{document}